\documentclass[a4paper,11pt]{article}
\usepackage{pos}
\usepackage{tikz}
\usetikzlibrary{shapes,arrows,positioning}

\usepackage{natbib}

\usepackage{lineno}

\title{Deployment of the IceCube Upgrade Camera System in the SPICEcore hole}
 \ShortTitle{SPICEcore hole Camera}

\author{The IceCube Collaboration \\{\normalsize \normalfont(a complete list of authors can be found at the end of the proceedings)}}

\emailAdd{ctoennis@icecube.wisc.edu}
\emailAdd{rott@physics.utah.edu}

\abstract{
IceCube is a cubic-kilometer scale neutrino telescope located at the geographic South Pole. The detector utilizes the extremely transparent Antarctic ice as a medium for detecting Cherenkov radiation from neutrino interactions. While the optical properties of the glacial ice are generally well modeled and understood, the uncertainties which remain are still the dominant source of systematic uncertainties for many IceCube analyses. A camera and LED system is being built for the IceCube Upgrade that will enable the observation of optical properties throughout the Upgrade array. The SPICEcore hole, a 1.7 km deep ice-core hole located near the IceCube detector, has given the opportunity to test the performance of the camera system ahead of the Upgrade construction. In this contribution, we present the results of the camera and LED system deployment during the 2019/2020 austral summer season as part of a SPICEcore luminescence logger system.

\vspace{4mm}
{\bfseries Corresponding authors:}
Christoph T\"onnis$^{1,2}$, Danim Kim$^{1*}$, Anna Pollmann$^{3}$, Carsten Rott$^{1,4}$

{$^{1}$ \itshape Department of Physics, Sungkyunkwan University, Suwon 16419, Korea}\\
{$^{2}$ \itshape Institute of Basic Science, Sungkyunkwan University, Suwon 16419, Korea}\\
{$^{3}$ \itshape Dept. of Physics, University of Wuppertal, D-42119 Wuppertal, Germany}\\
{$^{4}$ \itshape Department of Physics and Astronomy, University of Utah, Salt Lake City, UT 84112, USA}

$^*$ Presenter

\FullConference{37$^{\rm{th}}$ International Cosmic Ray Conference (ICRC 2021)\\
		July 12th -- 23rd, 2021\\
		Online -- Berlin, Germany}

}

\begin{document}
\maketitle

\section{Introduction}

The IceCube Neutrino Observatory is a cubic kilometer-scale Cherenkov radiation detector deployed in the ice near the geographic South Pole at depths between 1,450\,m and 2,450\,m. On the surface IceCube is complemented by a square kilometer large cosmic ray air shower detector. The main scientific goals of the experiment are to measure the high-energy astrophysical neutrino flux and to identify its sources. The IceCube collaboration reported the first observation of astrophysical neutrinos in 2013~\cite{HESE2013}. In 2018 the collaboration identified a flaring Blazar as a source of astrophysical neutrinos~\cite{TXS}. Other scientific goals of the detector include measurements of neutrino oscillation parameters, the cosmic ray flux, and the search for dark matter and other exotic particles. 

For the reconstruction of neutrino events in the IceCube detector, an accurate understanding of the optical properties of the Antarctic ice is crucial. To achieve this, various efforts have been made, in particular, using the LED flasher system on board of the optical modules of IceCube.  From these efforts, a model of the optical properties of the ice has been constructed~\cite{SPICE}. 

These optical properties still remain a dominant detector systematic uncertainty in many IceCube analyses. Based on calibration data collected with the IceCube LED flasher calibration system evidence for anisotropy in photon propagation was found, and a model based on variation of scattering as a function of photon direction was proposed~\cite{BFR2}. This was further substantiated using measurements with the IceCube dust logger~\cite{dustani}. Current models show that this is due to the South Pole Ice being a birefringent optical medium and take into account the shape and orientation of the ice microcrystals and the flow of the ice under its own weight~\cite{BFR1}. This anisotropy still remains to be studied in greater detail to aid in the development of ice models that incorporate this phenomenon accurately.

An upgrade comprised of seven densely instrumented strings in the centre of the active volume of IceCube is planned to enhance the detector. As part of this upgrade new types of Digital Optical Modules (DOMs) are going to be used, regularly spaced with a vertical separation of 3\,m between depths of 2160\,m and 2430\,m below the surface of the ice on each string. These new modules will carry novel calibration devices for the detector.

The IceCube upgrade camera~\cite{ICU_camera} is one of these calibration devices and designed to contribute to this ice model by measuring the scattering length of light in the ice using back-scattered light from an illumination system designed for the cameras. As a first test this camera has been deployed inside the IceCube luminescence logger~\cite{UV_logger} to measure the optical properties of ice surrounding a drillhole near the IceCube detector site.

\subsection{SPICEcore hole}

The SPICEcore hole was drilled during austral summer seasons of 2014/2015 and 2015/2016 near the Amundsen-Scott South Pole Station. The hole was drilled for glaciological, geological and climatological studies~\cite{SPICEcore_hole}. It is 1,751\,m deep and located about 1\,km away from the centre of the IceCube detector site. The instrumented volume of IceCube begins at 1,450\,m, but due to favorable tilt of the ice layers the relevant overlap with the hole is further extended by about 70\,m. The hole is being kept open by using an industrial anti-freeze agent, ESTISOL\textsuperscript{TM}-140, to allow for various scientific experiments~\cite{UV_logger, luminescence_logger}.

\section{Hardware}

\begin{figure}
    \centering
    \includegraphics[width=0.9\textwidth]{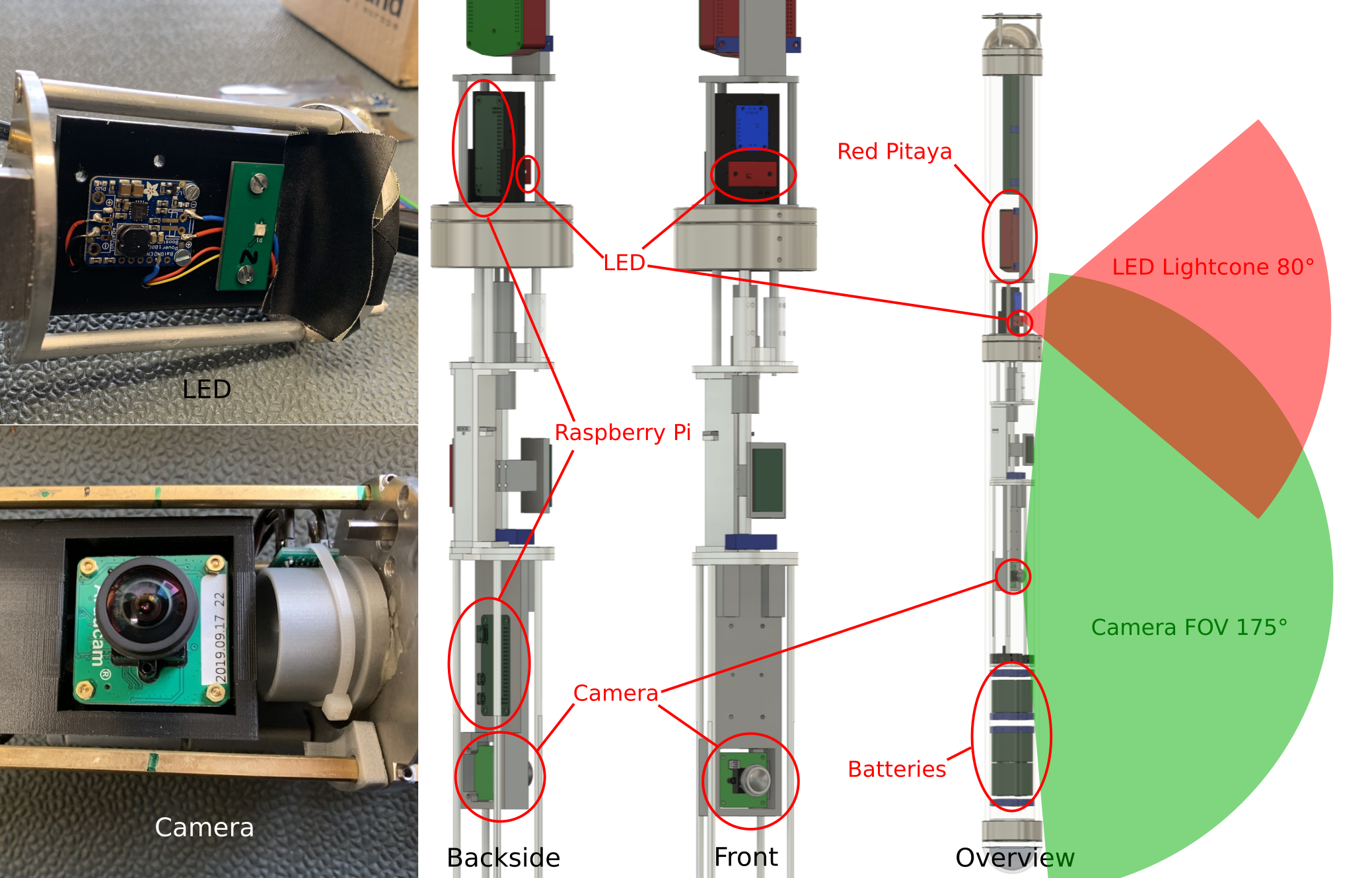}
    \caption{Photos and 3D diagrams of the camera and LED inside the luminescence logger from different viewing angles. The position of the camera and the LED board are highlighted. The field of view of the camera and the LED light cone are shown.}
    \label{fig:logger}
\end{figure}

One IceCube upgrade camera was integrated into the luminescence logger using a special holding structure replacing parts that were not needed for this measurement as shown in Figure \ref{fig:logger}. 

Close to the Red Pitaya, the main computer of the logger, an LED board with an SSL 80\,GB CS8PM1.13 LED from Oslon was installed. The board uses 1\,W of power generating 43\,lm of light with a dominant wavelength of 470\,nm. The cone of the light emitted has a full width at half maximum of 80 degrees. The LED is powered from a set of batteries located at the bottom of the logger.

When this LED emits light into the ice, as shown in Figure \ref{fig:logger} a fraction of the back-scattered light is detected by a camera situated in the lower half of the device. Images are captured continuously during the deployment. The scattering length can then be estimated based on the overall brightness of the images.

The camera is operated using a Raspberry Pi zero board that in turn is controlled by a Red Pitaya, an FPGA-based test and measurement board. The Pitaya was controllable via a RS 232 serial communication protocol using a graphical user interface from the surface during deployment. Commands to the Raspberry Pi are transferred via a serial peripheral interface using the secure shell (ssh) protocol.

The logger is equipped with a sensor chip comprising a magnetometer, an inclinometer and an accelerometer that was used to determine the orientation of the logger during deployment. The depth of the logger was measured by reading out the cable length automatically from the winch that was used to lower the logger into the ice from the surface.

The camera uses an IMX225 image sensor from SONY that is controlled by a MachX02 FPGA by Lattice semiconductor via a Inter Integrated Circuit (I2C) interface. There is a 8\,MB RAM on the camera that serves as a frame buffer that can store up to 3 images on the device before transferring data. The images captured by the camera have a resolution of 1312 by 979 pixels with a depth of 12 bits making a file size of 2.7\,MB per image.

\section{Deployment}

\begin{figure}
\begin{minipage}{0.22\textwidth}
    \centering
    \includegraphics[width=1.0\textwidth,keepaspectratio]{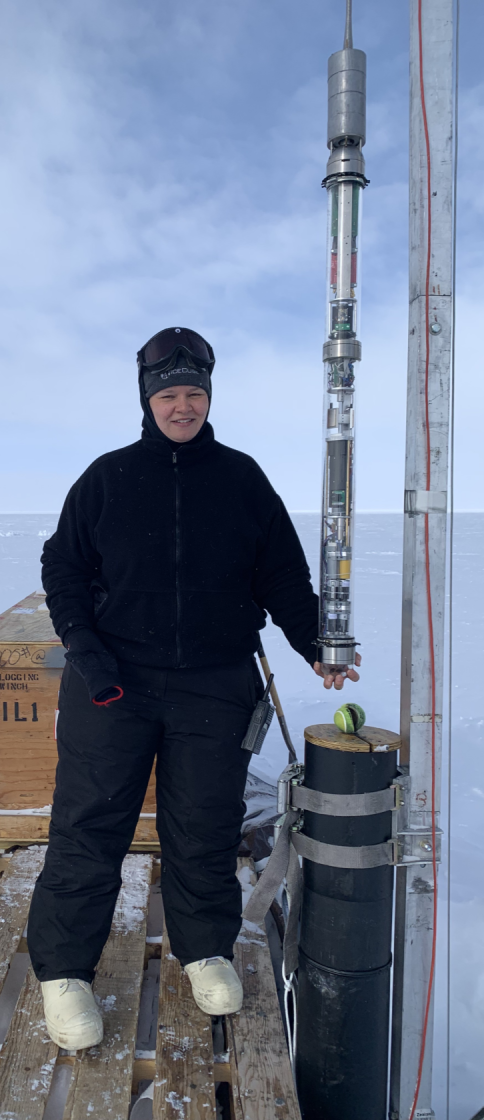}
\end{minipage}
\begin{minipage}{0.76\textwidth}
    \centering
    \includegraphics[width=1.0\textwidth,keepaspectratio]{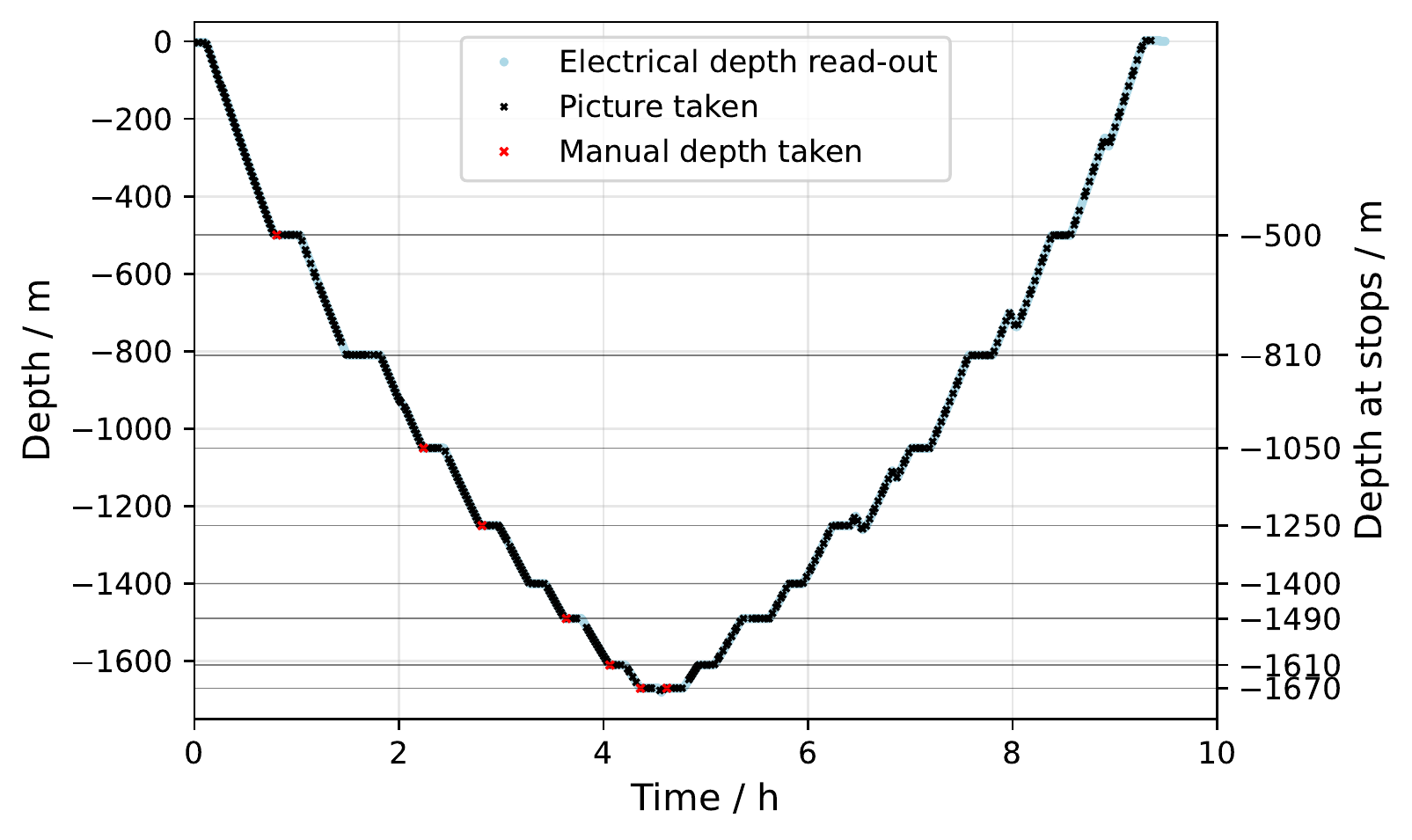}
\end{minipage}
\caption{Left: Photo taken at the SPICEcore hole with the logger prior to deployment. Right: The logger depth during deployment as a function of time. Green lines mark the stops during ascent and descent.}
    \label{fig:hole}
\end{figure}

\begin{figure}
    \centering
    \begin{minipage}{0.32\textwidth}
    \includegraphics[width=\textwidth]{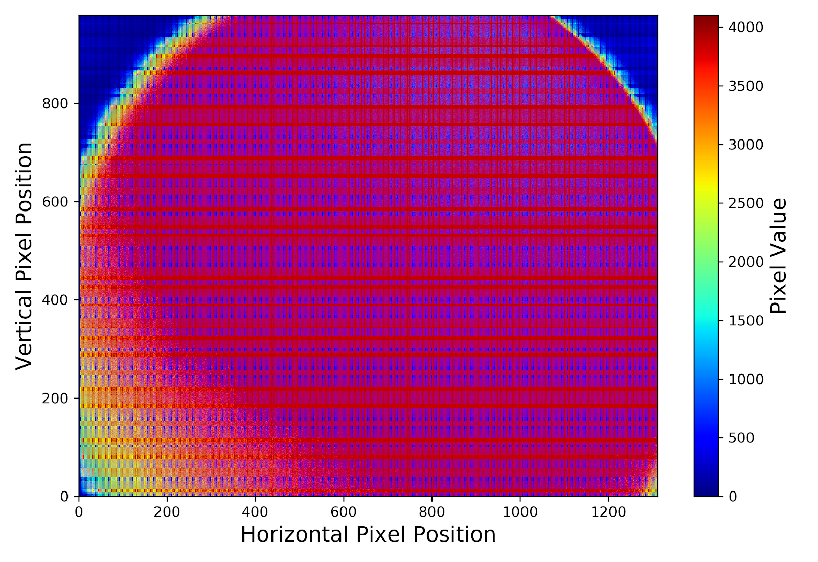}
    \end{minipage}
    \begin{minipage}{0.32\textwidth}
    \includegraphics[width=\textwidth]{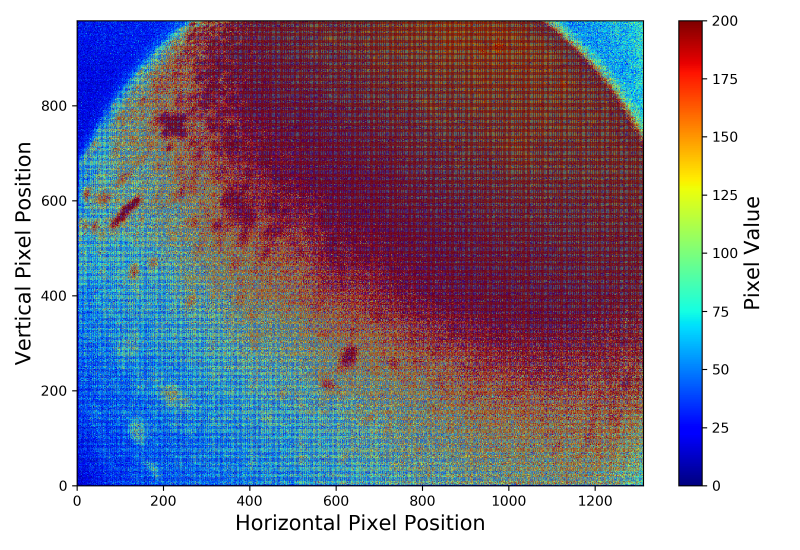}
    \end{minipage}
    \begin{minipage}{0.32\textwidth}
    \includegraphics[width=\textwidth]{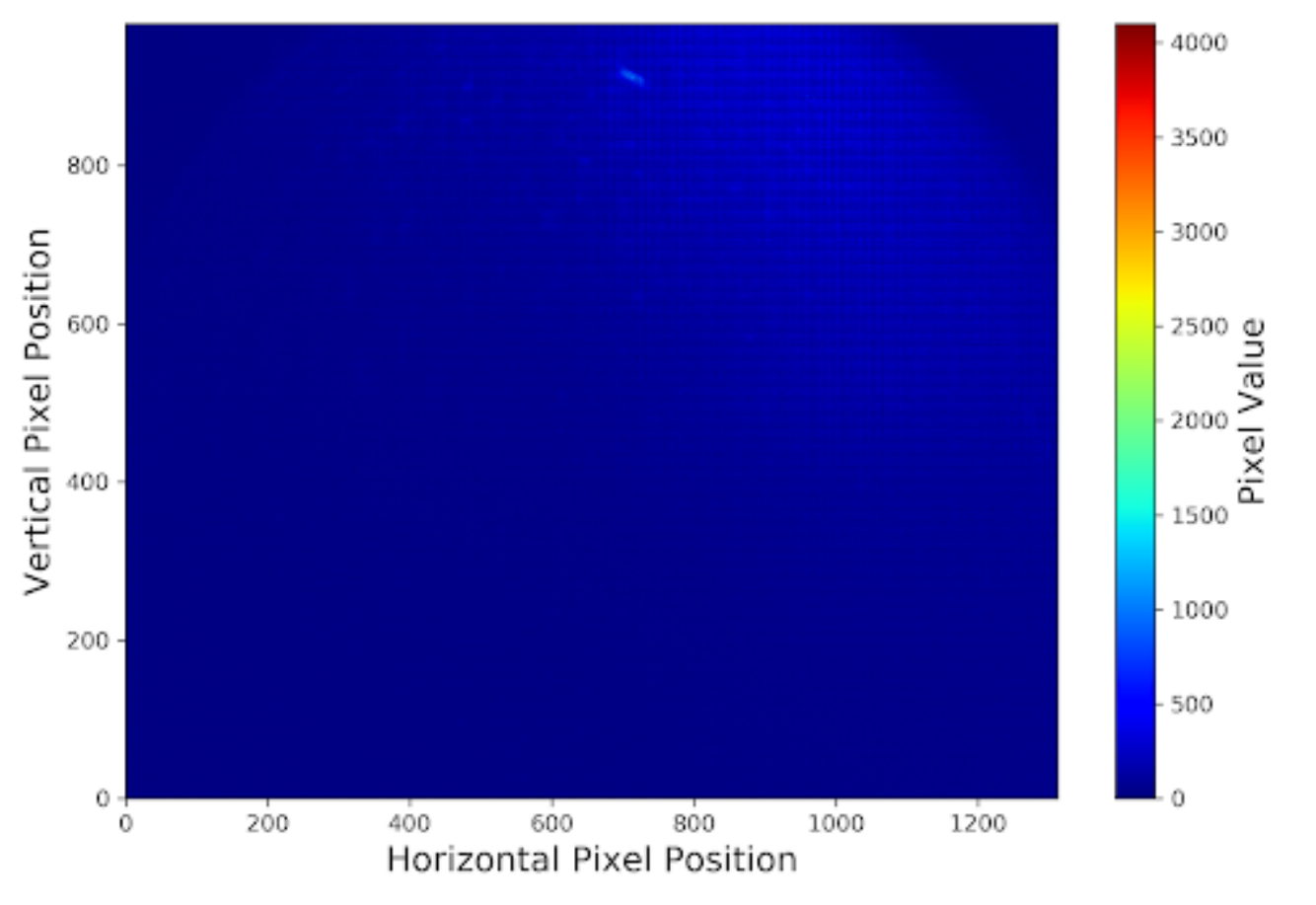}
    \end{minipage}
    \begin{minipage}{0.32\textwidth}
    \hspace{0.09\textwidth}
    \includegraphics[width=0.71\textwidth]{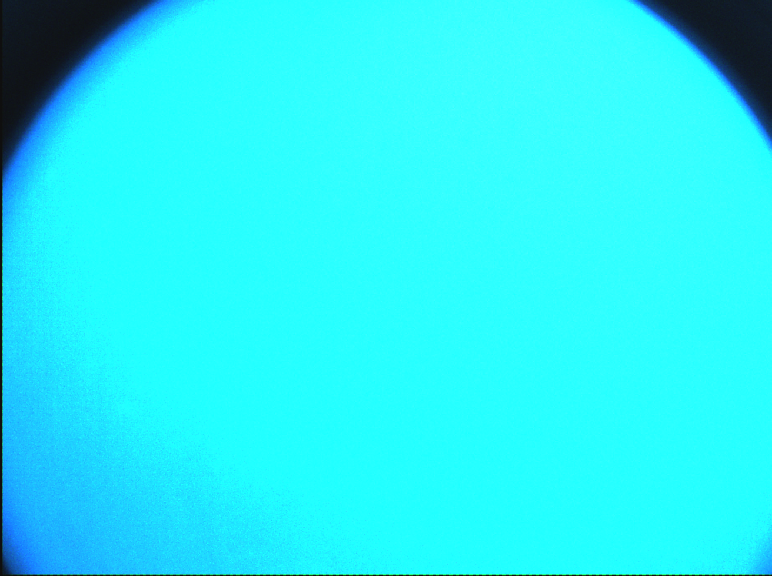}
        \end{minipage}
    \begin{minipage}{0.32\textwidth}
    \hspace{0.09\textwidth}
    \includegraphics[width=0.71\textwidth]{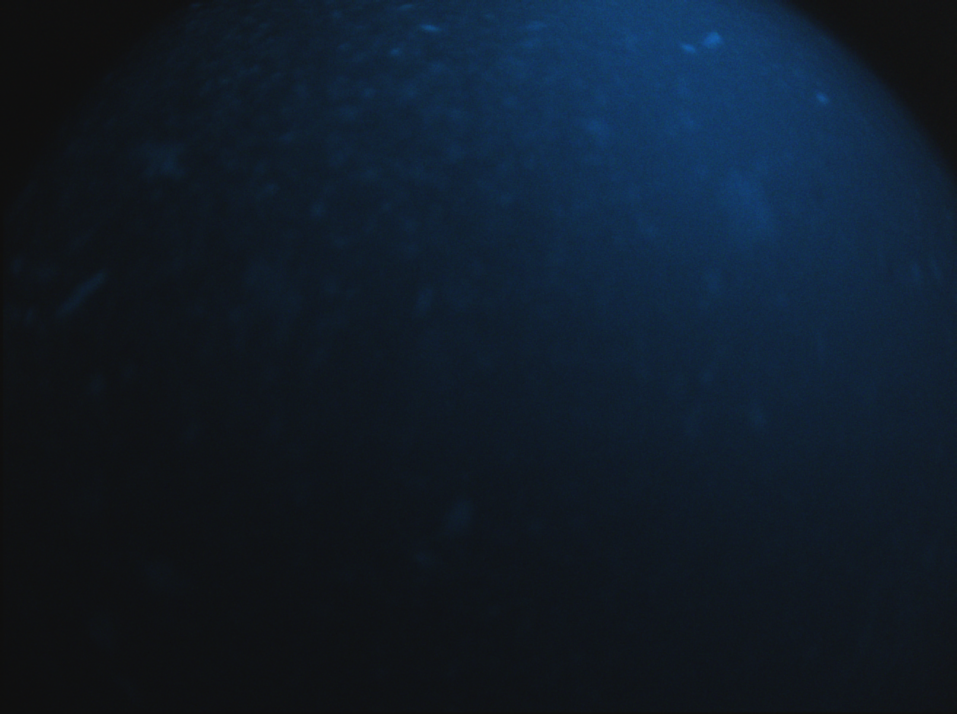}
    \end{minipage}
    \begin{minipage}{0.32\textwidth}
    \hspace{0.09\textwidth}
    \includegraphics[width=0.71\textwidth]{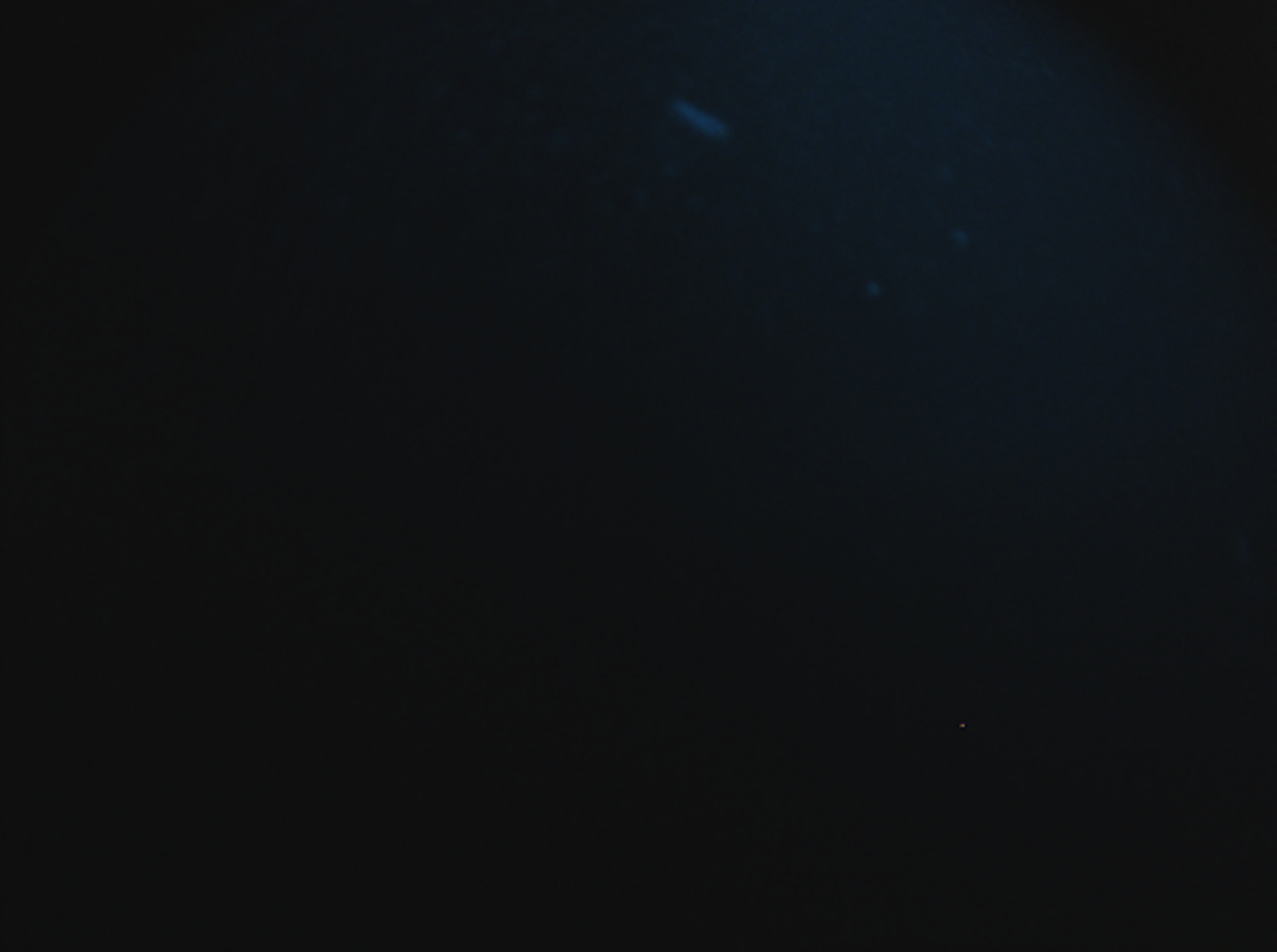}
    \end{minipage}
    \caption{Images captured with the camera at various depths. From left to right pictures taken at 810\,m, 1250\,m and 1610\,m are shown. The top row shows false color plots of the image data and the bottom row are true color images. All images use the same settings and while the LED was shining. The corners of the images are not illuminated due to the size of the lens and the image sensor. The images at 1250m show the flakes in the estisol very clearly.}
    \label{fig:photos}
\end{figure}

During the 2019/2020 winter season the luminescence logger~\cite{luminescence_logger} equipped with one IceCube Upgrade camera was deployed at the South Pole ice core hole. The logger was in the SPICEcore hole for a total of 10 hours on December 21st. The logger made stops at depths of 500~\,m, 810\,m, 1050\,m, 1400\,m, 1490\,m, 1610\,m and 1670\,m at both descent and ascent. 30 images were taken at each stop and continuously when the logger was moved. All images were captured with an exposure time of 3\,s and a gain of 0\,dB was used for images captured at or above 1500\,m and 24\,dB below. This was decided by the operator after checking live that the images are not too dim or too bright.  While descending the LED was enabled only during the stops to capture images and it was always enabled during ascent. Images taken during stops at descent are shown in Figure \ref{fig:photos} for different depths.

\begin{figure}
    \centering
    \includegraphics[width=\textwidth]{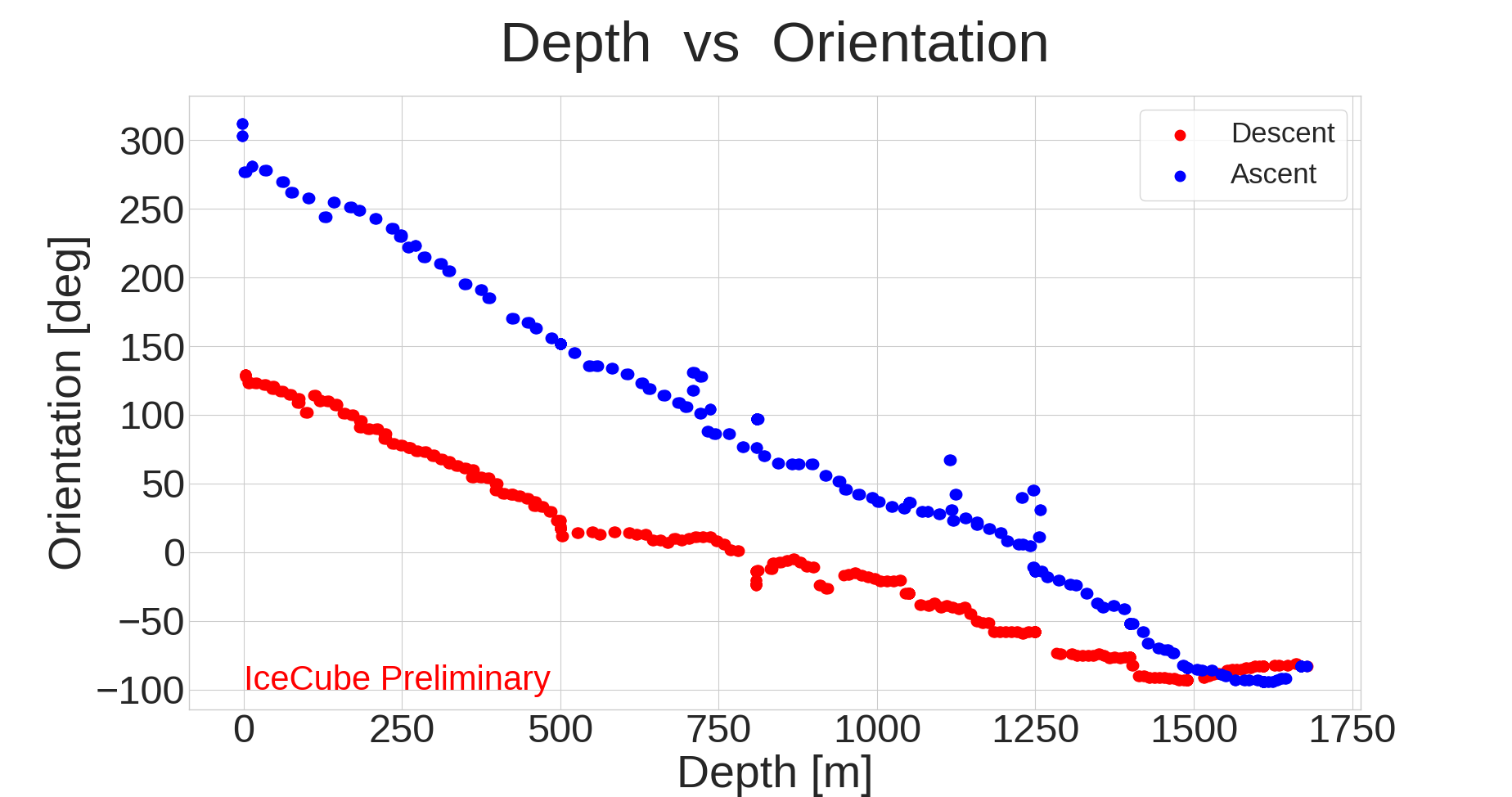}
    \caption{The orientation of the logger as a function of depth during deployment. 0 degrees were chosen arbitrarily and this plot serves to show the amount of rotation during deployment.}
    \label{fig:depth_angle}
\end{figure}

During ascent stops were made at the same depths and the LED was enabled continuously. Independent of the capture of images a log of the general logger operation, including logger orientation and depth was kept during ascent and descent. A plot of the orientation as a function of depth during ascent and descent can be seen in Figure \ref{fig:depth_angle}. During the descent the logger made a 180 degree revolution and during ascent turned in the other direction by approximately 360 degrees. In the Figure one can see the logger turning in place a little during the stops mentioned above.

\section{Analysis}

\begin{figure}
    \centering
    \includegraphics[width=\textwidth]{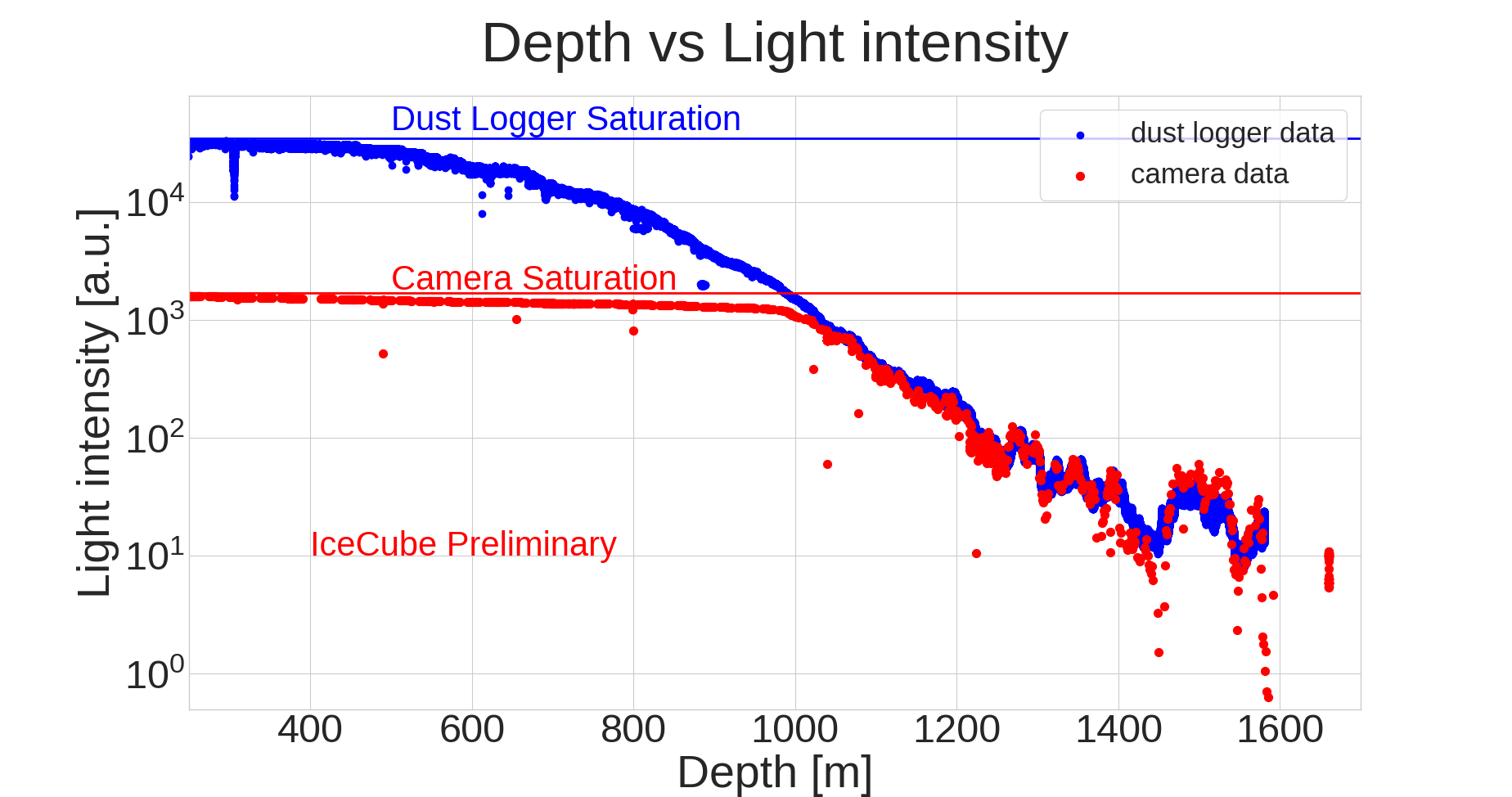}
    \caption{The dust logger and the camera brightness data as a function of depth. The camera data is scaled to fit to the dust logger data. Images captured between 1600\,m and 1670\,m were too dim to be used in the analysis.}
    \label{fig:depth_light}
\end{figure}

In order to show the capability of this type of camera system to measure scattering lengths in ice, the camera brightness data were compared to measurements conducted with the dust logger~\cite{dust_logger}. The dust logger is another logger deployed at the SPICEcore to measure the scattering length of light in the ice using a photomultiplier tube to measure the back-scattered light emitted by a laser.  A comparison of dust logger data with the image brightness from the camera as a function of depth is shown in Figure \ref{fig:depth_light}.

\begin{figure}
    \centering
    \includegraphics[width=\textwidth]{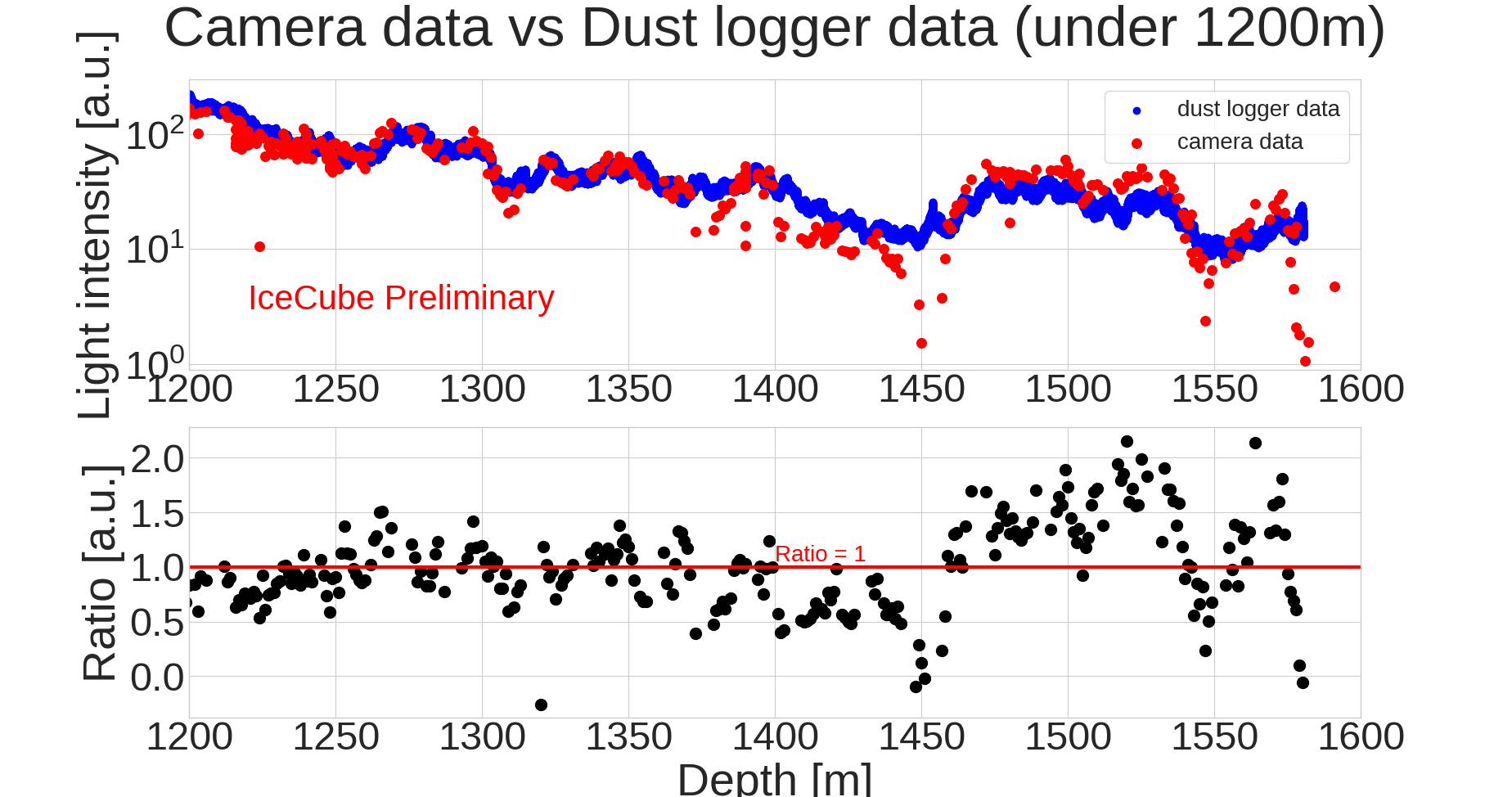}
    \caption{Similar to Fig. \ref{fig:depth_light} but zoomed in to the depth region between 1200m and 1600m. Comparison of the dust logger and the camera brightness data as a function of depth. In the lower part the ratio of camera and dust logger brightness is shown.}
    \label{fig:close_comparison}
\end{figure}

A zoomed-in version of this plot at deeper depths is shown in Figure \ref{fig:close_comparison}. The ratio between the two curves is nearly constant, in particular in the regions where the brightness is constant as, for instance, between 1470 and 1530\,m. 

\begin{figure}
    \centering
    \includegraphics[width=\textwidth]{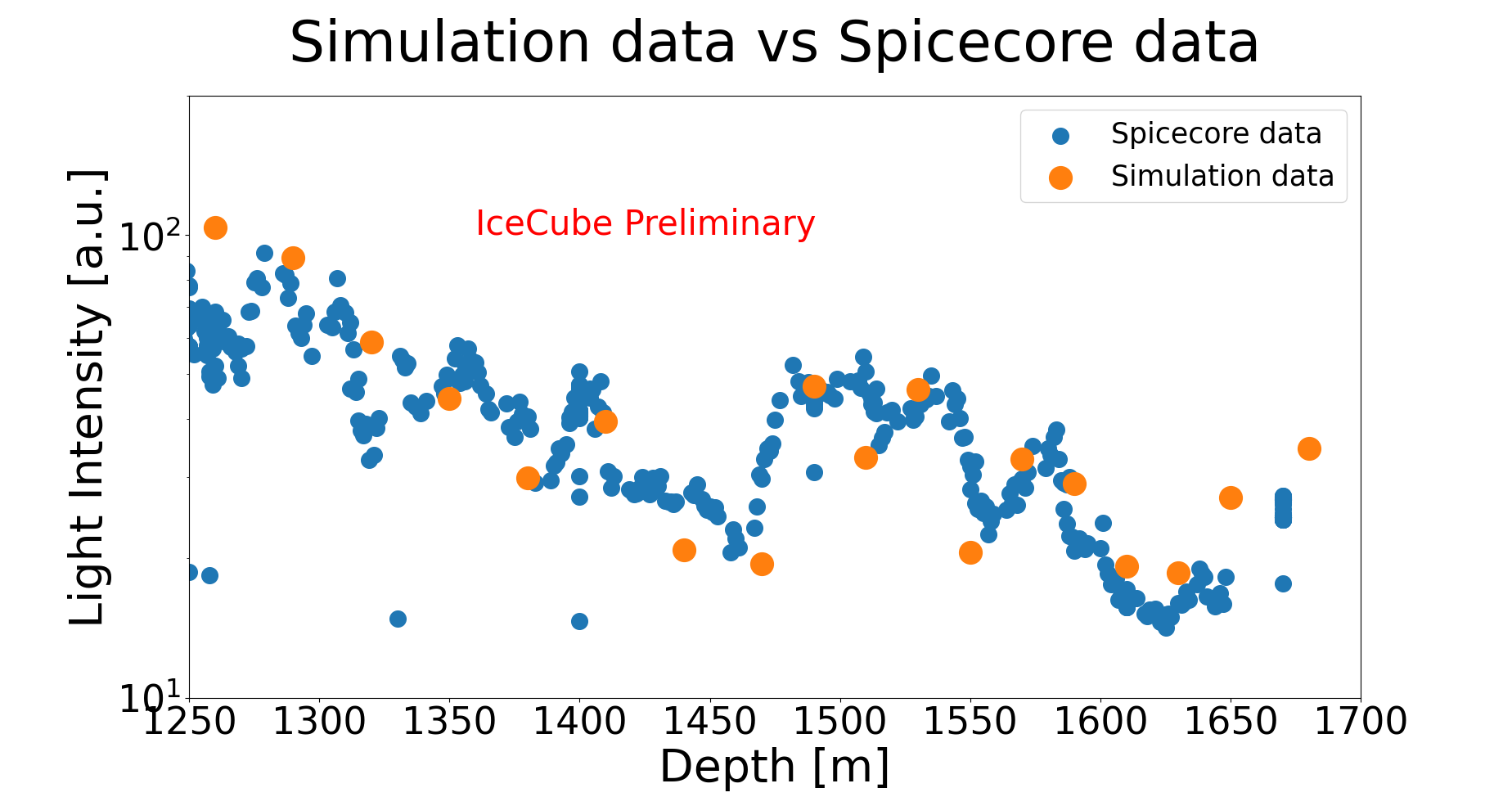}
    \caption{A comparison of simulated images generated with the PPC simulation package.The simulation was based on the current IceCube South Pole ice (SPICE) model~\cite{SPICE}. Depth values were adjusted to account for the tilt in the ice layers based on measurements with the dust logger at the IceCube site and the SPICEcore. There is an uncertainty of 30\,m to the depth values of the simulation due to the tilt of the ice layers between the SPICEcore hole and the IceCube site.}
    \label{fig:my_label}
\end{figure}

For further analysis simulated images were generated using the IceCube Photon Propagation Code (PPC) simulation~\cite{PPC}. The simulation used scattering length values from the IceCube ice model used in the IceCube neutrino event reconstruction~\cite{SPICE}. Simulations were then run for scattering length values evaluated at regular intervals in depth. The depth values in the SPICEcore hole and the IceCube detector were correlated using dust logger measurement data at each site. There is a strong correlation  between the simulated and the measured brightness indicating a good capability to measure the scattering length of the Antarctic ice using cameras as shown in Figure~\ref{fig:my_label}.

\section{Conclusion}

After a successful deployment of the IceCube upgrade camera system during the 2019/2020 austral summer in the SPICEcore hole a first analysis of the collected data has been conducted. The capability of the system to measure the optical properties of the Antarctic ice was demonstrated with a simple analysis method. A future analysis should be able to provide independent data on the scattering and absorption parameters of the ice that is complimentary to existing logger data.

\section*{Acknowledgement}
The authors would like to thank the SPICEcore collaboration for providing the borehole, the US Ice Drilling Program, the Antarctic Support Contractor and the NSF National Science Foundation for providing the equipment to perform the described measurement and for their support at South Pole.

\bibliographystyle{ICRC}

\bibliography{main.bib}

\clearpage
\section*{Full Author List: IceCube Collaboration}




\scriptsize
\noindent
R. Abbasi$^{17}$,
M. Ackermann$^{59}$,
J. Adams$^{18}$,
J. A. Aguilar$^{12}$,
M. Ahlers$^{22}$,
M. Ahrens$^{50}$,
C. Alispach$^{28}$,
A. A. Alves Jr.$^{31}$,
N. M. Amin$^{42}$,
R. An$^{14}$
% ...
\\

\noindent
% ...
$^{12}$ Universit{\'e} Libre de Bruxelles, Science Faculty CP230, B-1050 Brussels, Belgium \\
% ...
$^{14}$ Department of Physics and Laboratory for Particle Physics and Cosmology, Harvard University, Cambridge, MA 02138, USA \\
% ...
$^{17}$ Department of Physics, Loyola University Chicago, Chicago, IL 60660, USA \\
$^{18}$ Dept. of Physics and Astronomy, University of Canterbury, Private Bag 4800, Christchurch, New Zealand \\
% ...
$^{22}$ Niels Bohr Institute, University of Copenhagen, DK-2100 Copenhagen, Denmark \\
% ...
$^{28}$ D{\'e}partement de physique nucl{\'e}aire et corpusculaire, Universit{\'e} de Gen{\`e}ve, CH-1211 Gen{\`e}ve, Switzerland \\
% ...
$^{31}$ Karlsruhe Institute of Technology, Institute for Astroparticle Physics, D-76021 Karlsruhe, Germany  \\
% ...
$^{42}$ Bartol Research Institute and Dept. of Physics and Astronomy, University of Delaware, Newark, DE 19716, USA \\
% ...
$^{50}$ Oskar Klein Centre and Dept. of Physics, Stockholm University, SE-10691 Stockholm, Sweden \\
% ...
$^{59}$ DESY, D-15738 Zeuthen, Germany \\
% ...

\subsection*{Acknowledgements}

\noindent
USA {\textendash} U.S. National Science Foundation-Office of Polar Programs,
U.S. National Science Foundation-Physics Division,
U.S. National Science Foundation-EPSCoR,
Wisconsin Alumni Research Foundation,
Center for High Throughput Computing (CHTC) at the University of Wisconsin{\textendash}Madison,
Open Science Grid (OSG),
Extreme Science and Engineering Discovery Environment (XSEDE),
Frontera computing project at the Texas Advanced Computing Center,
U.S. Department of Energy-National Energy Research Scientific Computing Center,
Particle astrophysics research computing center at the University of Maryland,
Institute for Cyber-Enabled Research at Michigan State University,
and Astroparticle physics computational facility at Marquette University;
Belgium {\textendash} Funds for Scientific Research (FRS-FNRS and FWO),
FWO Odysseus and Big Science programmes,
and Belgian Federal Science Policy Office (Belspo);
Germany {\textendash} Bundesministerium f{\"u}r Bildung und Forschung (BMBF),
Deutsche Forschungsgemeinschaft (DFG),
Helmholtz Alliance for Astroparticle Physics (HAP),
Initiative and Networking Fund of the Helmholtz Association,
Deutsches Elektronen Synchrotron (DESY),
and High Performance Computing cluster of the RWTH Aachen;
Sweden {\textendash} Swedish Research Council,
Swedish Polar Research Secretariat,
Swedish National Infrastructure for Computing (SNIC),
and Knut and Alice Wallenberg Foundation;
Australia {\textendash} Australian Research Council;
Canada {\textendash} Natural Sciences and Engineering Research Council of Canada,
Calcul Qu{\'e}bec, Compute Ontario, Canada Foundation for Innovation, WestGrid, and Compute Canada;
Denmark {\textendash} Villum Fonden and Carlsberg Foundation;
New Zealand {\textendash} Marsden Fund;
Japan {\textendash} Japan Society for Promotion of Science (JSPS)
and Institute for Global Prominent Research (IGPR) of Chiba University;
Korea {\textendash} National Research Foundation of Korea (NRF);
Switzerland {\textendash} Swiss National Science Foundation (SNSF);
United Kingdom {\textendash} Department of Physics, University of Oxford.

\end{document}